\relax
\documentclass[reprint, aps, nofootinbib, superscriptaddress, longbibliography]{revtex4-1} 
\usepackage{amsmath, amssymb}
\usepackage{multirow}
\usepackage{verbatim}
\usepackage{xfrac}
\usepackage{graphicx}
\usepackage{dcolumn}
\usepackage{bm}
\usepackage{hyperref}
\usepackage{url}
\usepackage{listings}
\usepackage{booktabs}
\usepackage{tabularx}
\lstdefinestyle{custompython}{
  belowcaptionskip=1\baselineskip,
  breaklines=true,
  frame=L,
  xleftmargin=\parindent,
  language=Python,
  showstringspaces=false,
  basicstyle=\footnotesize\ttfamily,
}

\usepackage[usenames,dvipsnames,svgnames,table]{xcolor}

\begin{document}
\author{Allison C. Morgan}
 \email{allison.morgan@colorado.edu}
 \affiliation{Department of Computer Science, University of Colorado, Boulder, CO, USA}
\author{Dimitrios J. Economou}
 \email{dimitrios.economou@colorado.edu}
 \affiliation{Department of Computer Science, University of Colorado, Boulder, CO, USA}
\author{Samuel F. Way}
 \email{samuel.way@colorado.edu}
 \affiliation{Department of Computer Science, University of Colorado, Boulder, CO, USA}
\author{Aaron Clauset}
 \email{aaron.clauset@colorado.edu}
 \affiliation{Department of Computer Science, University of Colorado, Boulder, CO, USA}
 \affiliation{BioFrontiers Institute, University of Colorado, Boulder, CO, USA}
 \affiliation{Santa Fe Institute, Santa Fe, NM, USA}

\title{Prestige drives epistemic inequality in the diffusion of scientific ideas}\thanks{Published in EPJ Data Science {\bf 7}, 40 (2018)\\ \url{https://doi.org/10.1140/epjds/s13688-018-0166-4}.}

\begin{abstract}
The spread of ideas in the scientific community is often viewed as a competition, in which good ideas spread further because of greater intrinsic fitness, and publication venue and citation counts correlate with importance and impact. However, relatively little is known about how structural factors influence the spread of ideas, and specifically how where an idea originates might influence how it spreads. Here, we investigate the role of faculty hiring networks, which embody the set of researcher transitions from doctoral to faculty institutions, in shaping the spread of ideas in computer science, and the importance of where in the network an idea originates. We consider comprehensive data on the hiring events of 5032 faculty at all 205 Ph.D.-granting departments of computer science in the U.S. and Canada, and on the timing and titles of 200,476 associated publications. Analyzing five popular research topics, we show empirically that faculty hiring can and does facilitate the spread of ideas in science. Having established such a mechanism, we then analyze its potential consequences using epidemic models to simulate the generic spread of research ideas and quantify the impact of where an idea originates on its longterm diffusion across the network. We find that research from prestigious institutions spreads more quickly and completely than work of similar quality originating from less prestigious institutions. Our analyses establish the theoretical trade-offs between university prestige and the quality of ideas necessary for efficient circulation. Our results establish faculty hiring as an underlying mechanism that drives the persistent epistemic advantage observed for elite institutions, and provide a theoretical lower bound for the impact of structural inequality in shaping the spread of ideas in science.
\end{abstract}

\maketitle

\section{\label{sec:level1}Introduction}
A core principle of scientific progress is the free exchange of ideas, which enables the best ideas to spread throughout the scientific community. But, some ideas spread further than others, and these differences create a kind of epistemic inequality~\cite{wellmon:patronage}, in which some researchers and institutions are far more influential than others. These observed inequalities may reflect the impact of genuine differences in merit, or the importance of non-meritocratic factors associated with whom or where an idea originated, or both. Past studies of scholarly productivity show dramatic epistemic inequality: scientists at elite institutions produce the majority of research articles~\cite{wellmon:patronage}, play an outsized role in setting the pace and direction of scientific achievement~\cite{crane1965scientists,pelz76organizations,allison:department-effects,way:misleading}, and receive the majority of both professional awards and recognition~\cite{zuckerman1967nobel,allison1982cumulative,merton:matthew-effect,cole67rewards,moed2006bibliometric}. 

Such differences alone, however, are not clear evidence that epistemic inequality is driven by non-meritocratic social mechanisms, and there are very few data-driven tests for such mechanisms. As a result, it remains unknown how the spread of an idea may depend on where it originated in the scientific community. Moreover, if the point of origination does shape its fate within scientific discourse, what is the relationship between the idea's intrinsic fitness and the structural advantage afforded by the prestige of the origin? Progress on these questions would shed new light on systematic inequalities in scientific discourse and inform efforts to mitigate structural impediments to scientific progress. We also note that academia represents a kind of model system for studying socially-mediated information diffusion, as publications and institutions create a rich data ecosystem. As a result, insights on the spread of ideas in science may also yield new insights into other information diffusion settings, such as online social environments~\cite{cheng:cascades,bakshy:influence,watts:influence}. 

Past work on non-meritocratic factors that influence the spread of ideas in science has focused on two categories: institutional prestige and researcher prestige. Elite departments are known to provide resources that facilitate high rates of productivity and innovation~\cite{fox1983publication,smeby2005departmental}, including research funding, departmental staff, quality graduate students, and advanced facilities. Access to such resources can attract intrinsically talented researchers and foster environments that may naturally produce better ideas~\cite{longmcginnis81,hagstrom:prestige,kyvik1995big,stankovic2003recruitment}. 

Similarly, researcher influence itself can follow a cumulative advantage dynamic, called the ``Matthew Effect" in science, in which productivity and notoriety facilitate greater subsequent productivity and notoriety. As a result, well-known scientists tend to receive more credit than lesser-known scientists for work of comparable quality~\cite{merton:matthew-effect}. Furthermore, faculty in prestigious departments tend to be more visible to the research community \cite{cole1968visibility,weingart2005impact}, which can facilitate the spread of their ideas~\cite{cole:citation-prestige,aaltojarvi2008scientific,petersen2014reputation}. This greater visibility is often attributed to higher publication rates, greater representation in elite publication venues~\cite{van2005signals}, and greater engagement in informal scientific communication, e.g., circulating manuscripts and face-to-face communication~\cite{hagstrom:prestige}.

Here, we take a different approach, focusing instead on characterizing how faculty hiring drives epistemic inequality by determining which researchers are located at which institutions, and hence what ideas originate where. Faculty hiring can act as a transmission mechanism for the spread of research ideas, because researchers carry ideas that have been reinforced during their doctoral studies~\cite{petersen2007negotiating} to their faculty institution~\cite{clauset:hiring}. 
In this way, if a department becomes newly active in a particular research topic, it must have either hired as faculty a researcher who already works on that topic, or one of its existing faculty changed their research interests to align with the topic (e.g. via many other possible mechanisms such as conferences, social media, etc.). Hence, graduates who train under these faculty, who themselves go on to take faculty positions at still other institutions, and students of those faculty, etc., represent the continued spread of the idea, via faculty hiring alone, throughout the scientific community. From a historical perspective, the adoption of Feynman diagrams via the hiring of a small group of post-doctoral researchers from a single institution, represents an example of this mechanism~\cite{kaiser2005physics,rahmtin2017tracing}. 

To test the hypothesis that research ideas can spread to new universities through faculty hiring, we begin by analyzing the timing and topics of 200,476 computer science publications, and the hiring dates of 2,583 associated faculty.
Having found evidence to support this hypothesis, we then use comprehensive data on 5,032 faculty hires at the 205 Ph.D.-granting departments of computer science in the U.S. and Canada to construct a faculty hiring network that embodies the conduits along which ideas may flow among institutions. Using numerical simulations of simple epidemic models on this network, we quantify the impact on how far ideas of different inherent quality spread as a function of different originating institutions within the network. We find that ideas originating from prestigious universities spread faster and more completely than ideas from less prestigious universities, and we extract a generic ``exchange rate'' function that quantifies the tradeoff between increasing university prestige and decreasing quality of a research idea for generating an epidemic of a particular size.

The concept of a ``research idea" can span a diverse set of definitions, ranging from the development of a pioneering analysis technique or algorithm to the novel synthesis of previously disjoint observations. In Section \ref{sec:methods}, we evaluate hiring as one possible mechanism for the spread of ideas by identifying particular research topics via keywords in publication titles. For modeling purposes, in Section \ref{sec:models} we adopt a purposefully abstract definition of a ``research idea" as a meme with some intrinsic quality represented by the probability of transmission between two connected institutions. Accordingly, we consider the spreading of ideas at the level of institutions, where the adoption of an idea by a department is signaled by having at least one faculty member who's published research on that topic.
This construction is amenable to most reasonable concepts of a research idea and sheds light on the implicit tradeoffs between idea quality, network position, and the extent to which it spreads through the scientific community.

\section{\label{sec:data}Hiring and Publication Data}

Our study employs two comprehensive and complementary data sets. One contains detailed education and employment histories of  faculty at Ph.D.-granting computer science (CS) departments in the U.S.\ and Canada, along with data-driven estimates of each institution's ``prestige.'' The other contains the set of publications written by individual faculty who are listed in the first data set. Below we describe each in detail.

\begin{figure}[t]
\centering
\includegraphics[width=0.47\textwidth]{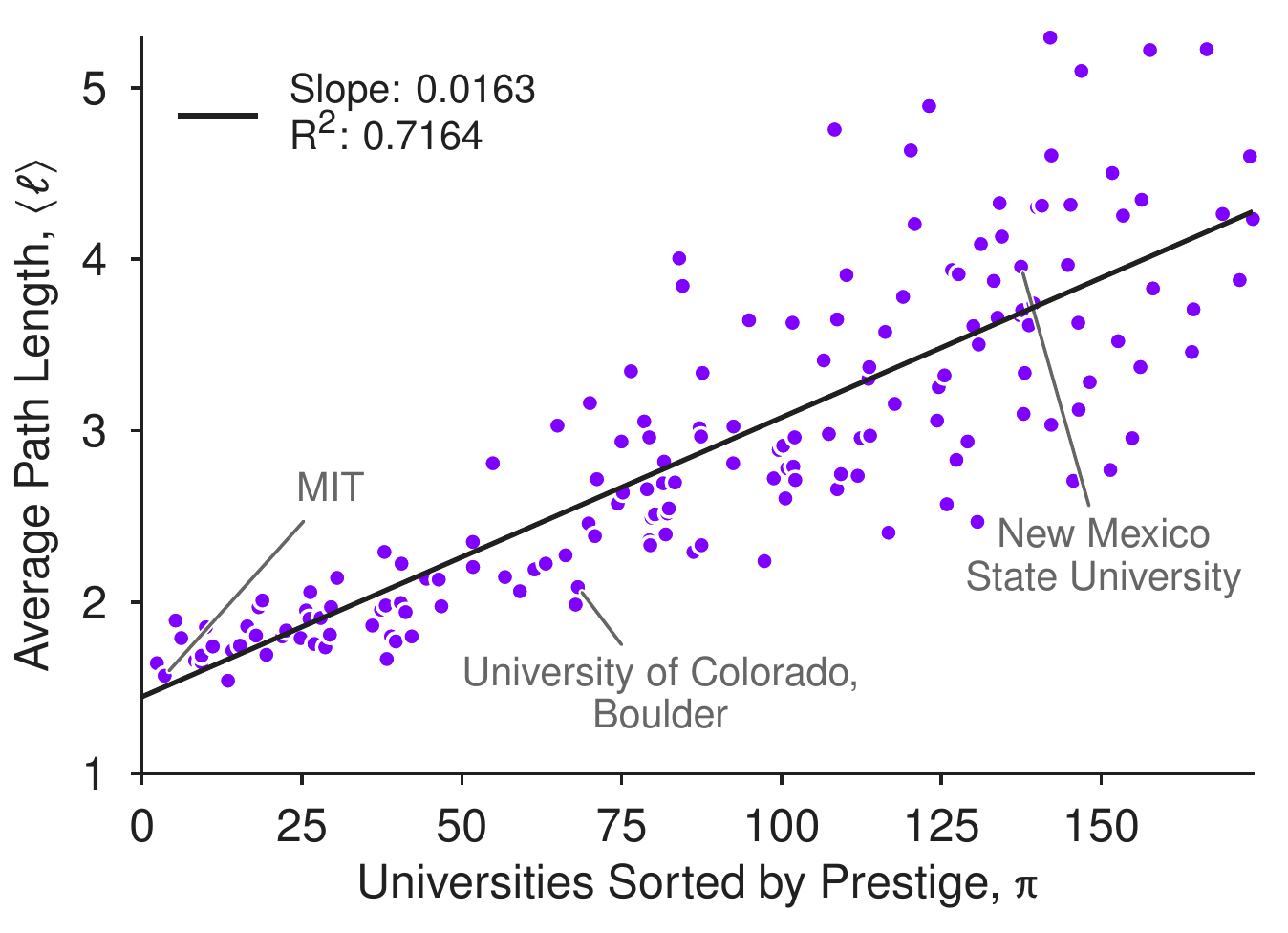}
\caption{Average length $\langle \ell \rangle$ of a geodesic path originating from a university with prestige $\pi$, in the strongly connected component of the computer science faculty hiring network, showing a strong linear correlation between a node's ``closeness'' centrality and its prestige}
\label{centrality}
\end{figure}

\subsection{\label{sec:network}Faculty hiring network}

We utilize an existing hand-collected data set of 5,032 tenured or tenure-track faculty from the set of all 205 Ph.D.-granting computer science departments in the U.S.\ and Canada~\cite{clauset:hiring}. From these data, we construct a multi-edge, directed faculty hiring network in which nodes represent universities, and an edge $(u, v)$ exists if a person received their Ph.D.\ from university $u$ and held a tenure-track position at university $v$ during the 2011--2012 academic year. Universities may have many edges between them, representing multiple researchers trained at $u$ who took a position at $v$. This network also contains self-loops, corresponding to individuals who received their Ph.D.\ at the same institution at which they hold a faculty position. We have omitted all non-Ph.D.-granting universities from our analysis, as well as faculty who received their Ph.D.\ from out-of-sample institutions.

\subsection{\label{sec:prestige}Departmental prestige}

Each institution in the data set is annotated by a data-driven estimate of its ``prestige'' within the faculty hiring system~\cite{clauset:hiring}, and we use this covariate to structure our investigation of how ideas spread differently depending on where the originate. Here, prestige measures a department's ability to place its graduates as faculty at other institutions, which has been shown to correlate with other departmental rankings (e.g., those compiled by U.S. News \& World Report and the National Research Council), but more accurately predicts faculty placement~\cite{clauset:hiring}.

Past research on visibility in science suggests that several institutional characteristics contribute to the success of individual researchers, in particular department size and prestige~\cite{cole1968visibility,aaltojarvi2008scientific}. Size is considered an ``almost necessary" condition for excellence~\cite{hagstrom:prestige} among academic departments, which require a minimum number of faculty in order to achieve sufficient breadth in research. However, other studies find that department size is only a weak predictor of success \cite{blackburn1978research,kyvik1995big} or has diminishing effect~\cite{jordan1988effects,dundar1998determinants} on the research output of faculty.

In contrast, departmental prestige is consistently an excellent predictor of faculty placement outcomes~\cite{clauset:hiring,way2016gender}, and hiring a graduate of $u$ as faculty at $v$ can be viewed as a kind of implicit endorsement of the perceived quality of $u$. Prestige also tends to correlate with department size and output~\cite{way:misleading}, but also allows small departments to have high placement power, or large departments to have low power. In addition to prestige, we also considered how other network-derived department characteristics correlated with spreading power, including eigenvector, in-degree (department size), out-degree (number of placed faculty), and closeness centrality scores. Of these, departmental prestige correlated most strongly, and hence we focus our investigation on how departmental prestige shapes the dissemination of research ideas.


%

Departmental prestige represents a node-level attribute extracted from the faculty hiring network, which is defined as a directed multigraph $G = (V, E)$, with $|V|=N$ nodes. A \emph{prestige hierarchy} is defined as a mapping $\pi : V \to [1,N]$, where $\pi_{i}$ is the prestige of node $i$, by convention $\pi_{i}=1$ is the highest prestige possible, and the number of ``rank violations'' is minimized. A rank violation is simply some edge $(u,v)\in E$ where the prestige of $v$ exceeds the prestige of $u$, i.e., the edge points ``up'' the hierarchy. In practice, however, there are many rankings $\pi$ with the same fraction of rank violations, and the prestige variable we use is the average prestige rank $\langle \pi_{i}\rangle$ over all minimum violation rankings~\cite{clauset:hiring}.

\begin{figure}[t!]
	\centering
\includegraphics[width=0.50\textwidth]{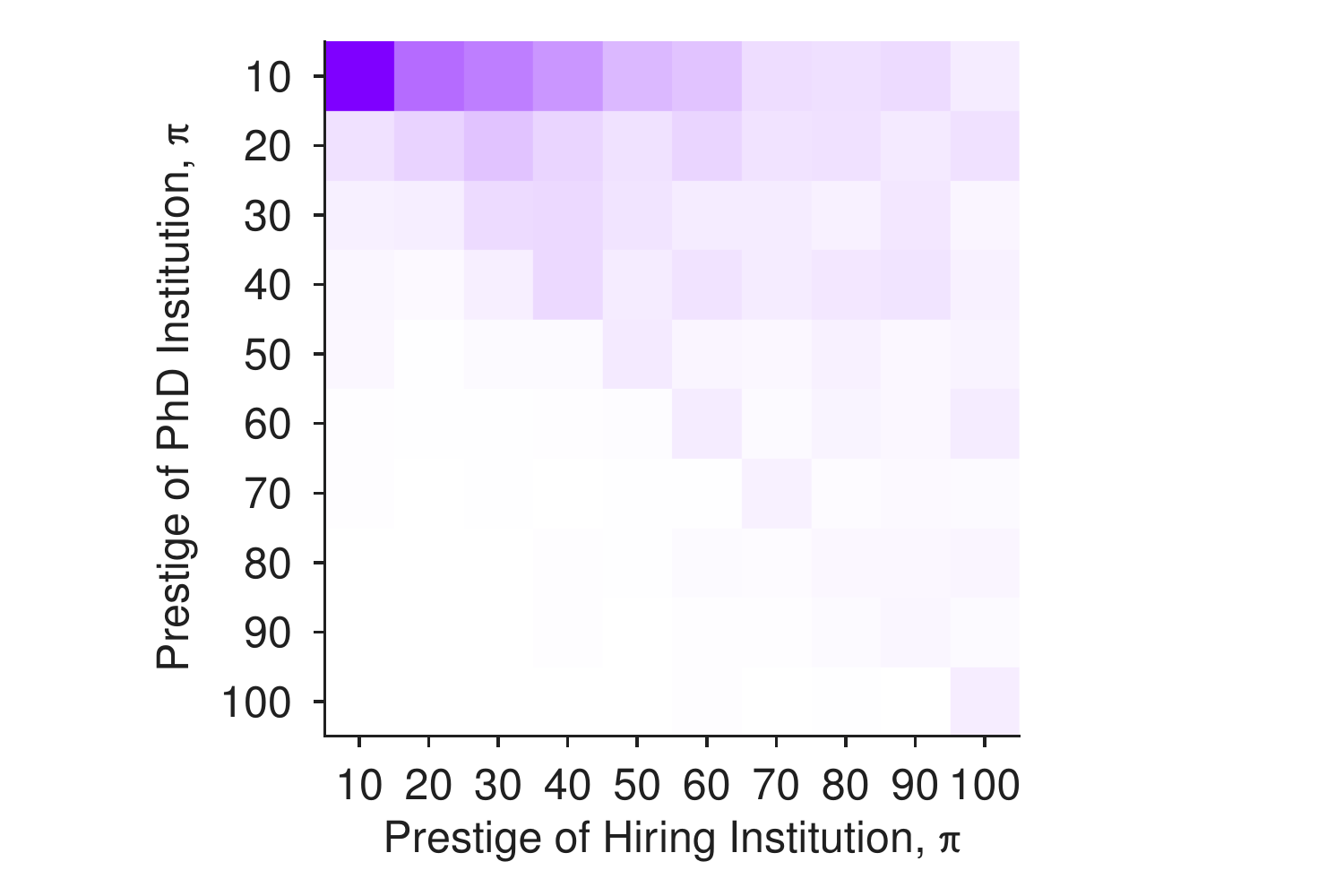}
  \caption{Coarsened adjacency matrix of the computer science faculty hiring network, sorted by prestige and aggregated into 10 groups. Shading is proportional to the density of edges between a pair of prestige deciles, and the strong upper triangle pattern indicates a strong prestige hierarchy.}
   \label{adjacency}
\end{figure}

There are three features of the CS faculty hiring network that are relevant for our study. First, the prestige hierarchy is steep, with only 12\% of CS faculty placing at institutions more prestigious than their doctorate. Second, it has a pronounced core-periphery structure~\cite{clauset:hiring}, in which prestige correlates with how ``close'' in the network a department is to other departments, as measured by the mean geodesic distance (Figure~\ref{centrality}). Third, there is enormous inequality in faculty production, and the number of placed faculty (out-degree) correlates with department prestige (Figure~\ref{adjacency}). This inequality is sufficiently extreme that only 18 (of 205) departments account for the doctorates of 50\% of all CS faculty in our data set~\cite{clauset:hiring}. Hence, in a practical sense, prestige drives faculty hiring.

\subsection{\label{sec:pubs}Faculty publications}
We also utilize an existing data set of papers authored by a subset of CS faculty listed in the faculty hiring network data set~\cite{way2016gender}. These data enable an empirical test of the mechanistic hypothesis that faculty hiring drives the spread of research ideas. Validating this mechanism provides the theoretical basis for our subsequent simulation-based investigation. 

In-sample faculty for this data set are those from the faculty hiring network for whom both the doctoral department is known and the department of the first assistant professor appointment is known, which is the primary transition for faculty hiring. 
For these faculty, publication records were obtained by manually linking faculty profiles to publication records on DBLP,
\footnote{See \url{http://dblp.uni-trier.de}} an online database spanning major computer science journals and conference proceedings. The result is a list of the timing and titles of 200,476 publications by 2,583 professors, which has previously been shown to be a representative sample~\cite{way2016gender,way:misleading}. 

\section{\label{sec:methods}Faculty Hiring and Epistemic Inequality}
The strong core-periphery structure of the faculty hiring network implies that, in terms of spreading dynamics, elite institutions have a structural advantage. However, investigating the consequences of this structure is only meaningful if scientific ideas can and do spread by way of faculty hiring. In this section, we construct a simple test to evaluate whether faculty hiring is a mechanism that shapes the spread of ideas in the academic computer science community. This test can be carried out for any research area with a well defined and specific set of associated terminology, and which is widely adopted across the community. Here, we apply the test to five well-known areas that satisfy these criteria: (i) ``deep learning,'' (ii) ``topic modeling,'' (iii) ``incremental computation,'' (iv) ``quantum computing,'' and (v) ``mechanism design.''

For each topic, a list of 7--56 keywords was generated manually for us by a set of at least two experts working within the corresponding field. Using the DBLP publication data for our in-sample CS faculty, we then extracted the associated set of publications for a topic using simple keyword matching on publication titles, with a manual verification step to guarantee each paper's topical relevance, yielding 1116, 217, 71, 167, and 122 publications respectively. Searching for words in titles will likely result in an under-classification of publications relevant to a research topic. For the measures of ideas transmitted via faculty hiring considered below, since we require that relevant research is carried out at their institution before and after their hiring, it is possible that we have classified events which should have been labeled as a transmission due to hiring as not. Given this approach, our measurement of research adoption via hiring is likely a conservative estimate or lower bound on the true number of such events.

\begin{figure}
\includegraphics[width=0.475\textwidth]{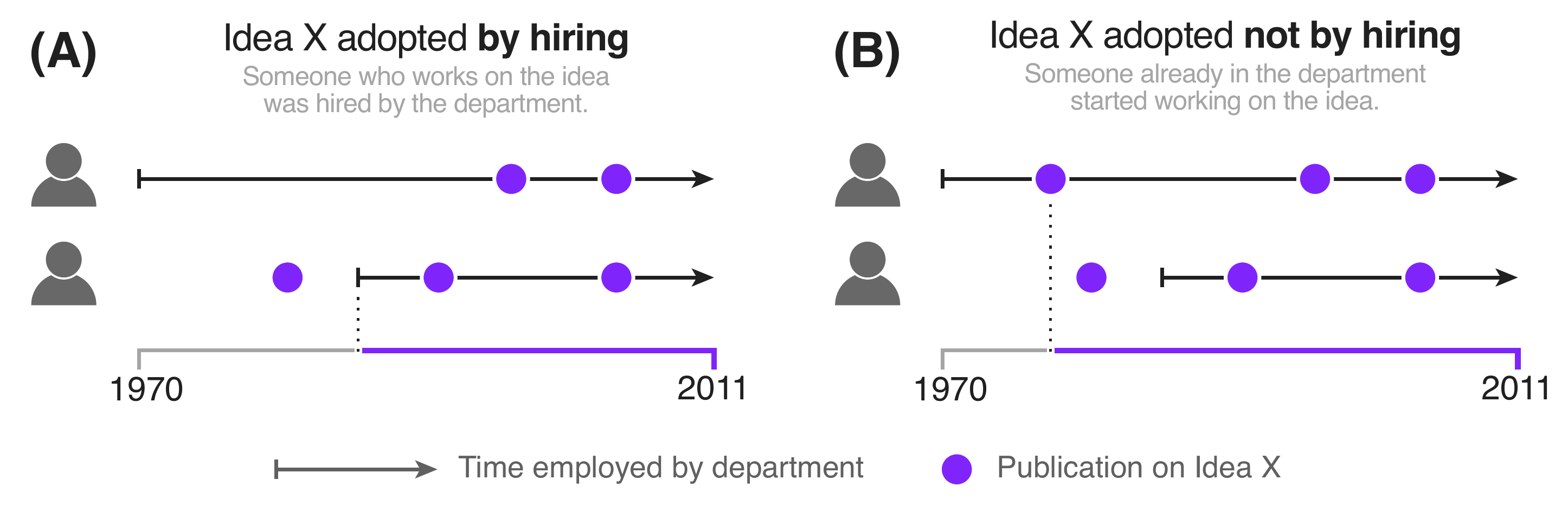}
\caption{If an idea X spreads to a department, the first person to work on X must either be (A) a newly hired faculty with prior work on X (hiring adoption), or (B) an existing faculty without prior work on X (non-hiring adoption). Black lines depict a faculty's time at a university, and purple dots signify relevant (on topic) publications.}
\label{example}
\end{figure}

Finally, for each faculty member j at each department i, we construct an indicator time series $f_{i,j}(t)=1$ if faculty $j$ published an on-topic paper in year $t$, and $f_{i,j}(t)=0$ if they did not. We then mark this time series with the year $t^{*}_{j}$ in which  $j$ was hired into the department.

For each department $i$, there are three scenarios for whether and how a topic X spreads to $i$:
\begin{enumerate}
\itemsep-0.1pt
\item (Null) X never spreads to department $i$, and hence for each faculty $j$ at department $i$, $f_{i,j}=0$. 
\item (Hiring) X spreads to $i$ (or, $i$ ``adopts'' X) by the hiring of new faculty $j$ who has previously published on X (Figure~\ref{example}A), i.e., a ``transmission'' of X from one department to another, carried by the new faculty $j$. In this case, no faculty at $i$ has published on X prior to the $t^{*}_{j}$, faculty $j$ has published on X prior to $t^{*}_{j}+2$, and $j$ publishes on X subsequently. The choice of allowing $j$'s ``prior'' work on X to occur up to 2 years after their faculty hiring event captures the fact that work carried out before being hired can take several years to be formally published.
\item (Non-hiring) X spreads to $i$ by one of their existing faculty publishing on X for the first time (Figure~\ref{example}B). In this case, some faculty $j$ at $i$ publishes on X prior to the hiring of any new faculty who have previously worked on X, without themselves representing scenario 2.
\end{enumerate}

\begin{figure}
\includegraphics[width=0.475\textwidth]{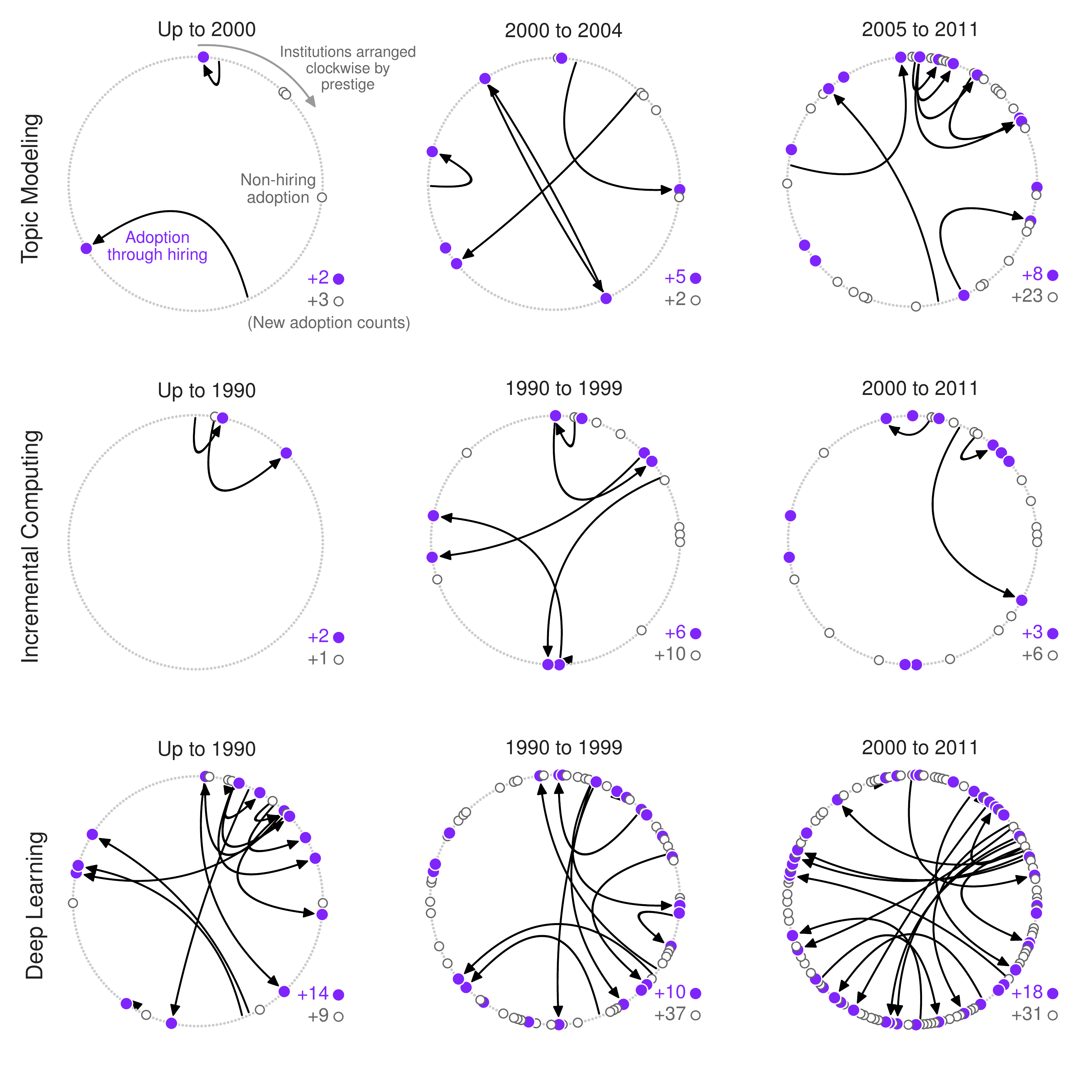}
\caption{Adoption events for the three research topics over time. Purple dots denote institutions who adopted an idea by hiring someone who studies that topic, and white dots represent institutions whose existing faculty began working on the topic. Arrows denote, for each time period, new transmissions, originating from the hired individual's doctoral location. All 205 institutions are arranged clockwise by prestige (descending), with the most prestigious department positioned at noon.}
\label{nettimes}
\end{figure}

Inspecting the time series of hiring events at all 205 universities, we recover a total of 241 spreading events for the five topics, each affecting between 11\% and 58\% of departments. Of these events, 88 (37\%) are due to transmissions of research ideas by way of hiring, and in 81\% of these cases, transmissions move via faculty from higher prestige universities to lower prestige universities (past studies show that only 9 to 14\% of faculty placements move faculty to a more prestigious university than their doctoral institution~\cite{clauset:hiring}). Figure~\ref{nettimes} illustrates these patterns by showing spreading events over time, for three of the topics.

Crucially, if faculty hiring shapes the spread of ideas, then a significant share of departments that ever adopt a topic X will have adopted it through faculty hiring (scenario 2). We test this hypothesis by constructing a specialized permutation test to assess the statistical significance of the empirically observed fraction of departments that have adopted a research idea via scenario 2, denoted $f_{\rm obs}$, and the expected fraction of such departments $f_{\rm exp}$. The test's null model is one in which the publication years for each faculty are fixed with their empirical values, but paper titles are drawn uniformly at random, without replacement, from the set of all titles. In this way, serial correlations in topics and temporal correlations with the hiring event are removed from each faculty. We then report empirical $p$-values~\cite{north:note} for the fraction of hiring-driven adoption events for each topic.

For all five research topics, the observed fraction of adoptions by hiring $f_{\rm obs}$ exceeds the expected fraction $f_{\rm exp}$ under the null model. However, the observed fraction was only statistically significant in four of five topics, topic modeling, incremental computing, quantum computing and mechanism design, while no significant difference was found for deep learning (Table~\ref{significance}).
\smallskip
\begin{table}[t!]
\caption{\textmd{Comparison between observed and expected adoptions via hiring for each research topic}}
\begin{tabular}{l|ccc}
topic X & $f_{\rm obs}$ & $f_{\rm exp}$ & $p$ \\
\hline
topic modeling & 0.35 & 0.23 & 0.01$\pm$0.01 \\
incremental computing & 0.39 & 0.20 & 0.01$\pm$0.01\\
deep learning & 0.35 & 0.34 & 0.34$\pm$0.01\\
quantum computing & 0.32 & 0.22 & 0.01$\pm$0.01\\
mechanism design & 0.48 & 0.21 & 0.01$\pm$0.01
\end{tabular}
\label{significance}
\end{table}
\smallskip

These results indicate that faculty hiring appears to act as a mechanism for 
the spread of ideas, with differential effects by topic, across the computer science community.
Faculty hiring has mostly clearly shaped the spread of topic modeling and incremental computing, and the lack of significance for deep learning is interesting. As previously discussed, this null result could be spurious, as the sampled nature of our data make it likely that we have underestimated the true share of departments that adopted a topic by hiring. However, it could also be related to the broad popularity of and interest in deep learning itself, which led to many more adoption events that were not driven by hiring. This case highlights the fact that faculty hiring may not play a statistically significant role in the spread of every research idea. At the same time, the other cases indicate that hiring does play a statistically significant role in others. The sample of research topics analyzed here should by no means be considered exhaustive, and, as such, we make no claims about the extent to which all research ideas spread via this mechanism. Instead, our results here establish that faculty hiring is a possible mechanism for the diffusion of ideas in academia, and we welcome future research to further explain which ideas spread by hiring and why.


\section{\label{sec:models} Prestige and the Diffusion of Ideas}

Having established empirically that faculty hiring itself plays a role in shaping the spread of real ideas across the scientific community, we now investigate the aggregate, system-level consequences of faculty hiring, and the links it creates between departments, on the spread of ideas, using numerical simulations. Our first model assumes faculty hiring is the sole mechanism by which research ideas spread throughout academia, and then we relax this assumption by allowing for diffusion via other mechanisms. This approach allows us to characterize how where an idea originates, and in particular the prestige of the originating department, shapes how broadly an idea may spread through faculty hiring, as a function of the idea's intrinsic quality. Hence, we quantify the degree to which ideas originating from more prestigious universities may spread more broadly than equally good ideas from less prestigious universities.

\subsection{Modeling the spread of ideas}
We model the spread of an idea across the CS faculty hiring network using a simple network model of information diffusion. Formally, this model is equivalent to an 
SI model in network epidemiology, repurposed here to model the spread of a meme~\cite{epidemicsurvey}.

In this model, nodes are in either a ``susceptible'' ($S$) or an ``infected'' ($I$) state; all nodes begin in state $S$; and, only the $S\to I$ state transition is allowed (no remission from infections). In the sense of ideas spreading, a department that adopts an idea (scenario 2 or 3 in Section~\ref{sec:methods}) undergoes the $S\to I$ transition. If some node $u$ undergoes the $S\to I$ transition, then in the next time step of the simulation, each of its susceptible neighbors independently undergoes the $S\to I$ transition with probability $p$,  where the chance for transmission of an idea across an edge is only allowed once (though multiple edges can exist between institutions). This probability quantifies the intrinsic quality or transmissibility of the idea, so that higher values of $p$ represent ideas that spread more easily. Finally, to initialize the simulation, a particular node $u$ is selected to undergo the $S\to I$ transition, and time then progresses until no new nodes transition to the $I$ state. 
This model assumes an independence between the prestige of node $u$ and the transmissibility of the idea $p$ originating at that node.
Additionally, more complicated epidemiological models are not considered here, but represent interesting directions for future work. For example, the SIS model allows a department to return to the $S$ state, e.g., by losing all its faculty who publish on a given topic. This model could be used to study the ebb and flow of interest in a topic across the network. 

For each department $u$ in the hiring network, we run a large number ($10,000$) of SI simulations with $u$ as the initial node, and we measure the mean epidemic size $Y$, i.e., the fraction of universities in state $I$ when the diffusion stops, and mean epidemic \emph{length} $L$, i.e., the number of time steps in the simulation. We then evaluate how these quantities covary with the prestige $\pi$ of the originating department, and the transmissibility of the idea $p$.

\subsection{Results}
\subsubsection{\label{sec:si}Simple SI Model}
\begin{figure}[t!]
	\centering
  \includegraphics[width=0.485\textwidth]{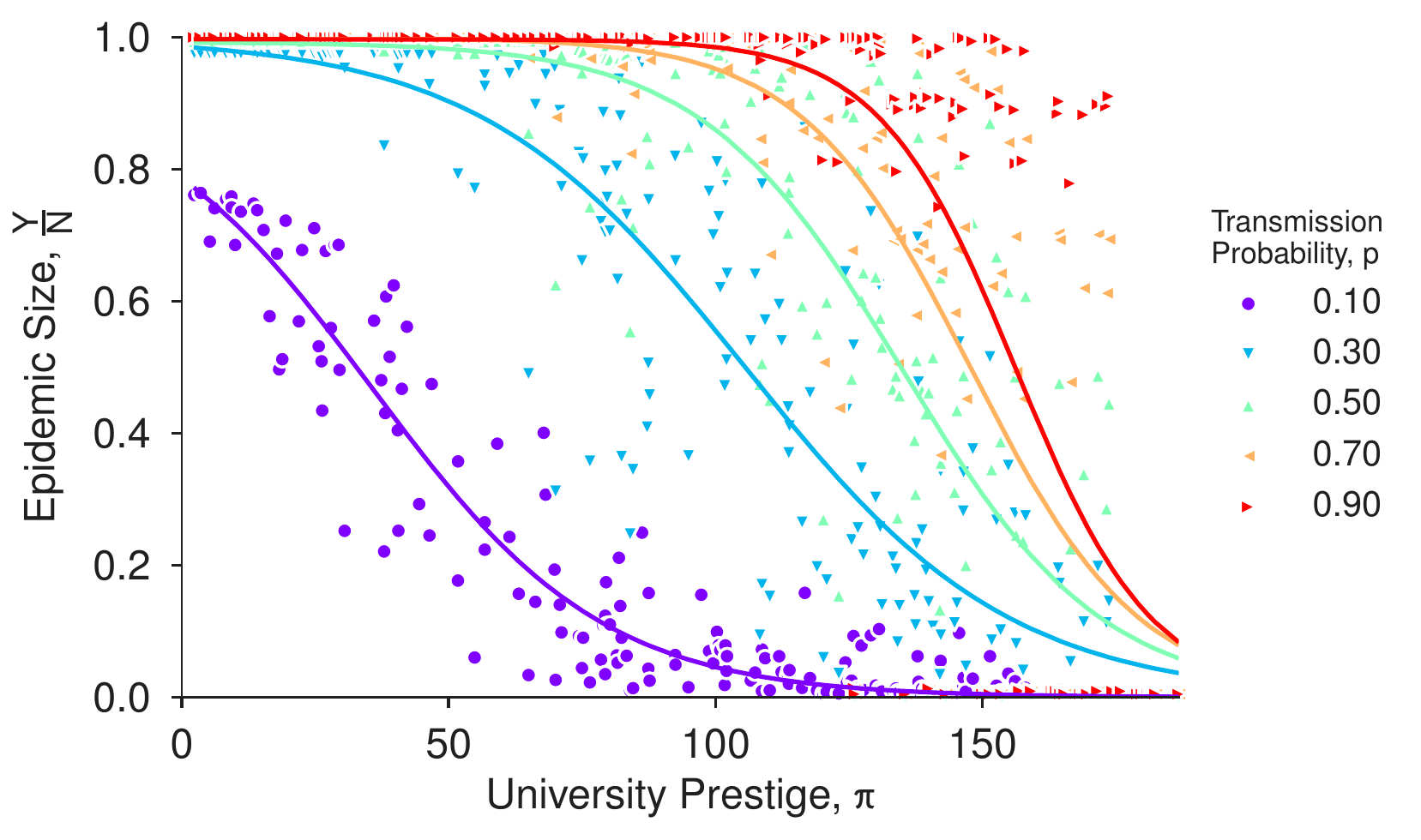}
  \caption{Normalized epidemic size $Y/N$ as a function of prestige of the originating university $\pi$. Each data point represents an average result over 1000 trials of the simulation. Colors correspond to different transmission probabilities $p$.}
  \label{SI-size}
\end{figure}

These information diffusion simulations show that ideas that originate at more prestigious universities tend to spread farther (larger epidemic size) than those originating at less prestigious universities, for ideas of similar quality (Figure~\ref{SI-size}).

\begin{figure}[t]
	\centering
  \includegraphics[width=0.465\textwidth]{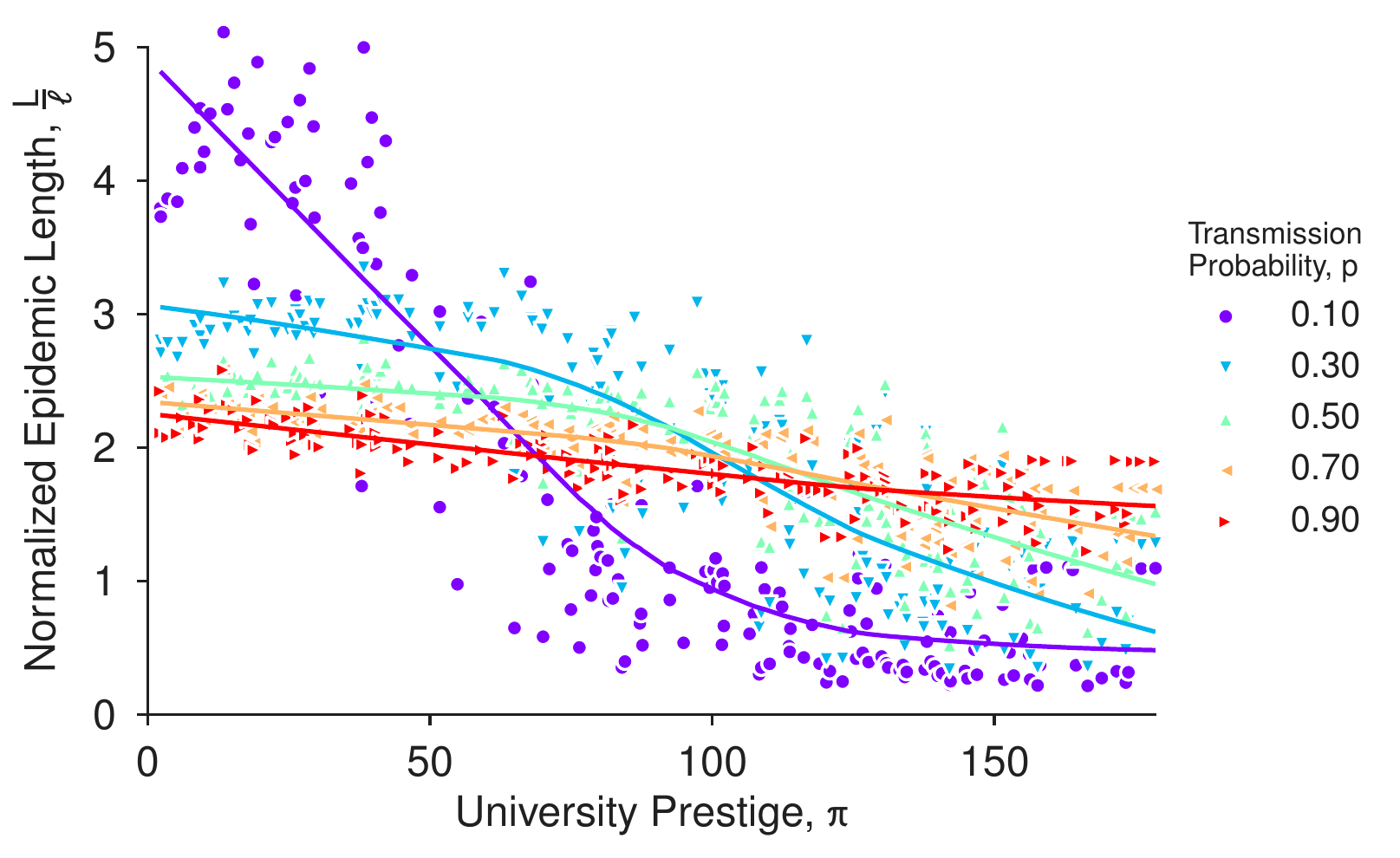}
  \caption{Normalized epidemic length $L/\ell$ as a function of prestige of the originating university $\pi$. Each data point represents an average result over 1000 trials. Each curve is colored corresponding to different transmission probabilities $p$ and fitted using a LOWESS curve.}
   \label{SI-length}
\end{figure}

This difference reveals a structural advantage that correlates with university prestige and its impact is most pronounced for lower-quality ideas. That is, lower-quality ideas (smaller $p$) that originate at more prestigious institutions will tend to spread farther than comparable ideas (same $p$) that originate at less prestigious institutions. Accordingly, increasing $p$ has a more dramatic effect on increasing the corresponding epidemic size produced by lower prestige institutions. In other words, our simulations suggest that high-quality ideas will tend to spread throughout the network regardless of where they begin (although at different timescales, Figure~\ref{SI-length}). But, the structural advantage of higher prestige tends to enhance the circulation of lower-quality ideas, and is likely related to the way prestige correlates with with increased network centrality and more faculty alumni.
Notably, these simulation results corroborate past empirical studies of the effects of prestige on researchers' citations and visibility~\cite{cole:citation-prestige,cole1968visibility}. 

To model the effect of prestige $\pi$ on the size $Y$ of the resulting epidemic, we fit logistic curves to the results:
\begin{align}
  \frac{Y}{N} = y_\text{max} \left /\, 1 + \textrm{e}^{-k (\pi - \pi_\text{mid})} \right. \enspace , \nonumber
\end{align}
where $y_\text{max}$ is the upper bound of the size, $k$ is the growth rate, and $\pi_\text{mid}$ is the symmetric inflection point. The good visual agreement between a logistic growth curve and our simulation results (Figure~\ref{SI-size}) suggests that for a particular idea fitness $p$, there exists a range of prestige values within which linear increases in prestige result in exponential increases in epidemic size. For smaller values of $p$, this range is concentrated among the most prestigious universities, reflecting their structural advantage. As $p$ increases, the range shifts progressively toward lower-prestige universities. In other words, for linear increases in $p$, we observe non-linear epidemic sizes.

We also find that prestige shapes how long ideas tend to circulate in the network, as measured by the length $L$ of the epidemic, normalized by the average length of a geodesic path $\ell$ from the originating university (Figure~\ref{SI-length}). This ratio $L/\ell$ quantifies the degree to which an idea circulates beyond or below the shortest-path percolation. For high-quality ideas (larger $p$), we find that the epidemic length $L$ tends to be similar regardless of where an idea originates, although there is a slight positive correlation with prestige. However, lower-quality ideas from higher-prestige universities circulate much longer than if they originate from lower-prestige universities, again illustrating the structural advantage that prestige affords in the diffusion of ideas when we consider faculty hiring as the only mechanism by which ideas spread.

\subsubsection{\label{sec:jump}SI Model with Jumps}
We now relax the importance of faculty hiring by introducing a stochastic ``jump'' into the transmission model, which models the aggregate effect of other spreading mechanisms---word-of-mouth, professional meetings, reading the literature, social media, etc.

Because faculty hiring tends to be highly selective on the prestige of hiring and placing institutions~\cite{clauset:hiring}, some universities are disconnected from large sections of the faculty hiring network. However, ideas that originate at these peripheral universities should still have some chance to spread through means other than faculty hiring or the communication conduits created by those relationships. To capture this effect, in the lifetime of an epidemic, each university $u$ that has made the $S\to I$ transition, in addition to its faculty hiring transmissions, will also transmit the idea to exactly one university, selected uniformly at random from $u$'s set of unreachable nodes, with ``jump'' probability $q$. This process mirrors the ``teleportation'' probability of random walkers in the PageRank algorithm~\cite{page:rank}.

\begin{figure}[t!]
	\centering
  \includegraphics[width=0.475\textwidth]{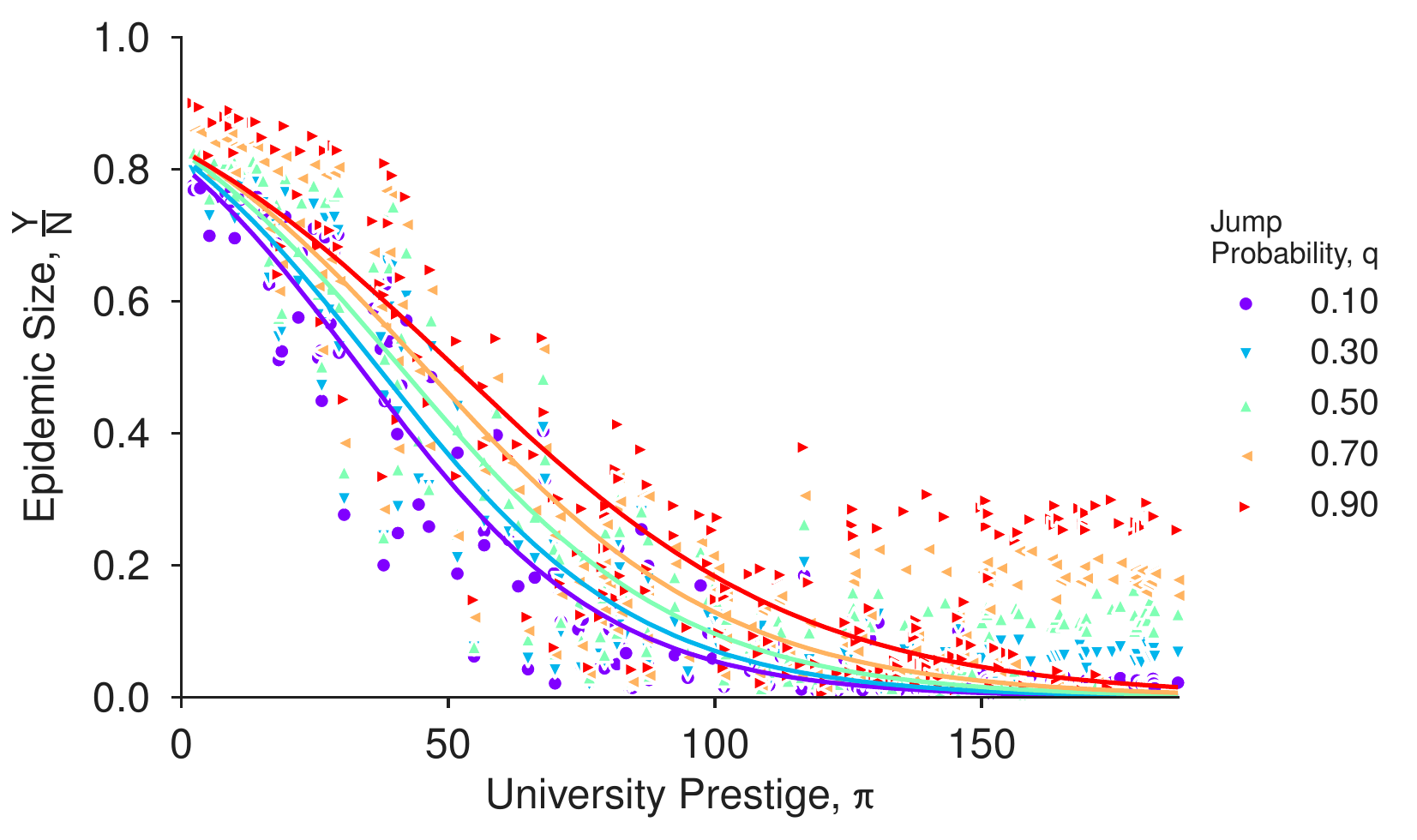}
  \caption{Normalized epidemic size $Y/N$ as a function of prestige of the originating university $\pi$, allowing for a single jump to a disconnected node. Transmission probability is held constant at $p = 0.1$. Each data point represents an average over 500 trials. Colors correspond to different jump probabilities $q$.}
   \label{SI-random-size}
\end{figure}

This variation of our information diffusion simulation shows that increasing the likelihood of this non-hiring transmission modestly improves the spread of ideas originating from the lowest prestige universities (Figure~\ref{SI-random-size}), as these universities now have some chance of transmitting an idea to a more central institution. Even very high jump probabilities, however, do not mitigate the strong structural advantage in spreading that the highest-prestige universities exhibit. Similarly, $q$ has only a marginal impact on the epidemic size produced by the highest prestige institutions, whose ideas already tend to spread widely across the network.

\begin{figure*}[t]
	\centering
  \includegraphics[width=0.9\textwidth]{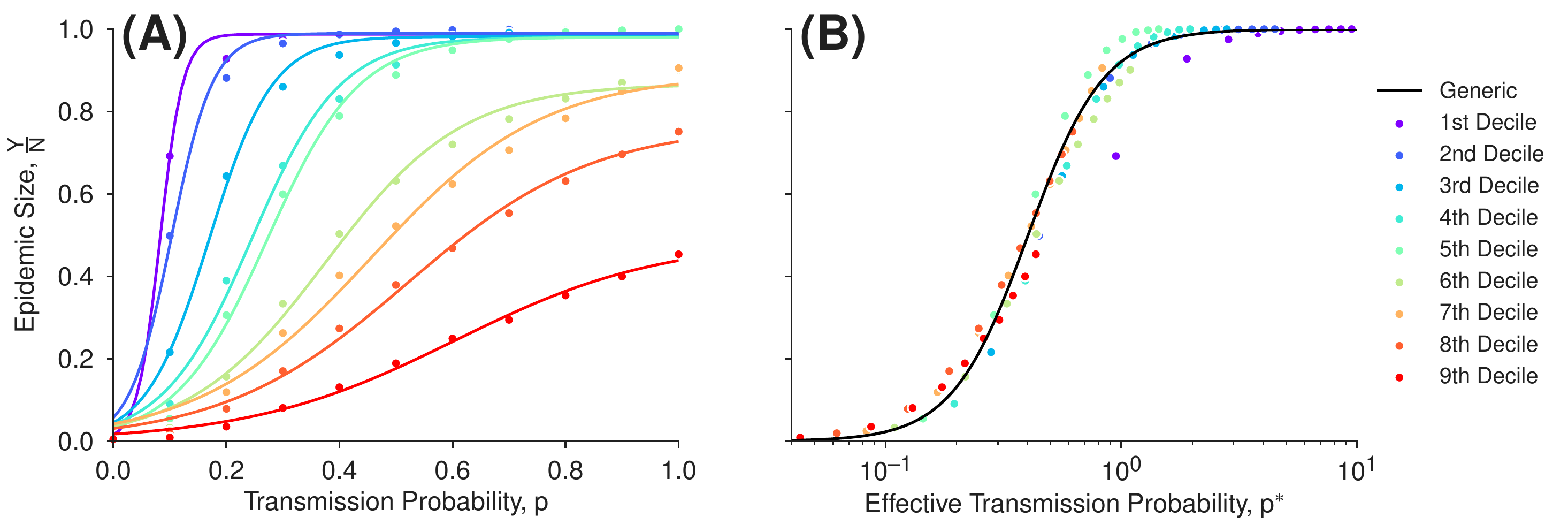}
  \caption{(A) Normalized epidemic size $Y/N$ as a function of transmission probability $p$ for each university averaged across the universities in a prestige decile. (B) Epidemic sizes for normalized transmission probabilities. Each data point represents an average across all universities in a decile, and across the 1000 trials for each university, transmission probability pair.}
  \label{SI-infection-size}
\end{figure*}

\subsubsection{\label{sec:infection}A Generic Tradeoff Between Prestige and Idea Fitness}

Knowing that prestige exerts such a strong influence on the spread of ideas across the network, we now consider whether there exists a generic relationship between the prestige of the originating university and the quality of the idea it is spreading. In this way, we aim to quantify the tradeoff between these two variables by asking: For a given epidemic size, how much must $p$ increase to compensate for a decrease in $\pi$?

To begin, we stratify institutions into decile groups according to their prestige. We then compute the average epidemic size $Y/N$ among universities in each decile, as a function of transmission probability $p$, and fit logistic functions to these data. This analysis reveals that ideas originating from the lowest-prestige universities, even if they are of the highest quality, are unlikely to spread to the whole network (red line, Figure~\ref{SI-infection-size}A), and again reinforces the substantial structural advantage afforded to more prestigious universities. As a result, less prestigious universities face substantial structural barriers in the spread of their original ideas, independent of their quality, which may play a role in persistent epistemic inequality and the dominance of elite universities in the pace and direction of scientific progress.

To quantify the precise relationship among prestige $\pi$, idea quality $p$, and epidemic size $Y$, we use a technique from statistical physics called a ``data collapse'' to extract a generic functional form. When a set of curves are parameterized special cases of a more general function, the generic function can be identified and estimated by ``rescaling'' the individual curves so that they collapse onto each other~\cite{newman1999monte}. To obtain this function, we rescale the decile curves in Figure~\ref{SI-infection-size}A using an \textit{ansatz} that relates $p$ and $d$, the decile of prestige, i.e., 0.1 for the top 10\% most prestigious universities, 0.2 for the next 10\%, etc.:
\begin{align}
p^{*} = -p \left /\, \log(1-d) \right. \enspace . \nonumber 
\end{align}
This ansatz converts $p$ and $d$ into an ``effective'' transmission probability $p^{*}$, and its form illustrates the exponential rescaling effect of prestige (via the decile variable $d$ now) on the raw transmission probability $p$. Hence, as the prestige of the originating university decreases, in order to produce an epidemic of equivalent size, the transmission probability of the idea must increase at an exponential rate to compensate.

Replotting the the epidemic size data as a function of the effective transmission probability produces the data collapse (Figure~\ref{SI-infection-size}B), and confirms the existence of a generic functional relationship, in which epidemic size varies as a function of $p$ and $d$ alone, via $p^{*}$:
\begin{align}
\frac{Y}{N} = \left[1 + \textrm{e}^{r(k + \log p^{*} )}\right]^{-1} \enspace , \nonumber 
\end{align}
where $r$ and $k$ are constants of best fit. Under this function, a choice of prestige decile $d$ that originates an idea and a choice of idea quality $p$ would allow us to roughly predict the fraction of the network the idea will eventually reach.

\section{\label{sec:conclusion}Conclusion}

Past studies of scholarly productivity and affiliation suggest that researchers at elite institutions play an outsized role in driving the pace and direction of scientific progress \cite{crane1965scientists,pelz76organizations,allison:department-effects,way:misleading,zuckerman1967nobel,allison1982cumulative,merton:matthew-effect,cole67rewards,moed2006bibliometric}. Here, using comprehensive data on faculty hiring events in the field of computer science, we investigated the consequences of a university's prestige on the diffusion of ideas it originates. Using epidemic models to simulate the spread of ideas across the faculty hiring network, we find that ideas originating from more prestigious universities produce larger epidemic sizes and longer epidemic lengths. Consequently, ideas starting in the network periphery (i.e., at less prestigious universities) must be much higher in quality to have similar success as lower quality ideas originating in the core (more prestigious universities). These findings suggest that idea dissemination within academia is not meritocratic, even when the assessment of the idea's quality (transmission probability $p$) is entirely objective.

While these results may appear intuitive, 
our study provides a detailed and quantitative characterization of the theoretical consequences of institutional prestige on the spread of ideas across the scientific community. These measurements build upon the notion that research ideas spread throughout academia by way of faculty hiring, either through the direct transfer of researchers working on a particular topic or by the lines of communication created between placing and hiring institutions~\cite{cole1968visibility,allison:department-effects}. We tested the hypothesis that faculty hiring acts as a mechanism for the spread of ideas by carefully cross-inspecting DBLP publication data and faculty hiring events, showing that indeed, faculty hiring plays a statistically significant role in driving the spread of some ideas. Specifically, the spread of the research topics incremental computation, topic modeling, quantum computing, and mechanism design was significantly driven by faculty hiring events in our network. The same could not be said for deep learning, however, which may suggest that deep learning is a less specialized, possibly less well-defined, research topic. Alternatively, deep learning may simply represent a particularly high-quality idea, whose adoption was both widespread and rapid (see Figure~\ref{nettimes}). 

Our investigation of these five areas of research was limited by matching keywords to titles of publications, and for computer science faculty only. Analysis of full-text articles or abstracts would facilitate more precise detection of publications relevant to particular research areas~\cite{gerow2018measuring}. Along these lines, a more detailed analysis of how robust the faculty hiring mechanism is under a more specific, or more broad, definition of a research area is an interesting and important direction of research. Future work should consider extending the analyses performed here to other departments where faculty hiring network data are available. Subsequently, our analyses of idea diffusion suppose that faculty hiring provides the primary conduit for the spread of research and models all other modes of diffusion using a small, uniform jump probability that connects all universities. To the extent that these other modes of diffusion are structured and can be measured, future work should consider modeling their effects directly to provide more realistic estimates of idea diffusion.

Additionally our work 
focuses on faculty hiring as a mechanism for the spread of ideas throughout academia.
Certainly, other mechanisms exist that influence the dissemination of ideas, including those mediated by the scientific literature and its underlying citation network. Our analysis of the random jump model helps explain the transmission of ideas under transmission mechanisms independent of prestige. If we believe that many other mechanisms that drive the spread of ideas are strongly correlated with prestige, as is the case for citation networks~\cite{cole1968visibility,crane1965scientists,moed2006bibliometric,helmreich1980making}, then despite the fact that these mechanisms are different from the one we are testing, the inclusion of their effects in our analysis might only slightly mitigate the structural advantage we observe.

Our results suggest that researchers at prestigious institutions benefit substantially from a structural advantage that allows their ideas to more easily spread throughout the network of institutions, and consequentially, impact the discourse of science. 
This advantage presumes that ideas spread according to a purely meritocratic notion of idea quality and that ideas of high quality can originate from any institution. If it is instead the case that the quality of an idea is correlated with its origination (i.e., high quality ideas are more likely to come from prestigious institutions) then the quality of an idea would act as a confounding factor to the faculty hiring mechanism. Producing an objective, empirical measurement of an idea's quality is difficult and would require, for example, an assertion of the relative worth of advancements in theory versus methodology, which remains an open problem. Nevertheless, past research supports the existence of a ``halo effect" in science~\cite{merton:matthew-effect,astin81,cutright2003pursuit}, whereby ideas are perceived as being of higher quality if they originate from prestigious institutions and researchers. As such, our results may indicate only a lower bound for the actual advantage of that elite universities enjoy. Future studies should consider modeling non-meritocratic factors such as the halo effect in addition to the purely meritocratic effects analyzed in our study.

A difficult question left unanswered by this work is what, if anything, should be done about the impact of non-meritocratic social mechanisms on epistemic inequality in scientific discourse. Our results indicate that faculty hiring is one social mechanism that drives this inequality, and past work has established that university prestige drives faculty hiring~\cite{clauset:hiring}. Hence, if 
differences in faculty placement rates across universities remain unchanged (i.e., institutions continue to hire faculty with doctorates from a small number of elite departments), then the current highly differential spread of ideas 
is unlikely to change on its own.

Somewhat more optimistically, our results show 
that while lower quality ideas will tend to be overshadowed by comparable ideas from more prestigious institutions, high quality ideas circulate widely, regardless of their origin. Epistemic inequality, then, could be mitigated by incentivizing researchers to produce a smaller number of higher quality studies over a larger number of incremental, low-quality studies. Additionally, there may be opportunities to mitigate epistemic inequality through new technologies and careful experimentation~\cite{clauset2017data}. For example, the adoption of double-blind review processes~\cite{goues2017effectiveness} and the practice of posting early manuscripts online may facilitate the visibility of high quality ideas from less prestigious universities. However, these online tools may also amplify academia's existing inequalities~\cite{beel2009google}. Continued experimentation of this form will be important for monitoring the effects of policy on epistemic inequality in science. We look forward to more work on understanding the mechanisms that create and maintain epistemic inequality, and innovative ideas to promote the free circulation of good ideas.

\begin{acknowledgements}
The authors thank Daniel Larremore, Chad Wellmon, Andrew Piper, Jon Kleinberg, Bailey Fosdick, and Jerry Senderson for helpful conversations. Authors ACM, SFW, and AC were supported by National Science Foundation Award SMA 1633791. ACM was also supported by a National Science Foundation Graduate Research Fellowship Award DGE 1650115. The dataset and code supporting the conclusions of this article are available in the \texttt{allisonmorgan/epistemic\_inequality} repository, \url{https://github.com/allisonmorgan/epistemic\_inequality}.
\end{acknowledgements}

%

%

\end{document}